\documentclass[%
 aip,
 amsmath,amssymb,
 reprint,%
]{revtex4-1}

\usepackage{graphicx}

\usepackage{dcolumn}
\usepackage{bm}

\usepackage[utf8]{inputenc}
\usepackage[T1]{fontenc}
\usepackage{mathptmx}
\usepackage{etoolbox}
\usepackage{xcolor}
\usepackage{textgreek}
\usepackage{hyperref}
\usepackage{cleveref}
\crefname{chapter}{Chapter}{Chapters}
\crefname{section}{Sect.}{Sects.}
\crefname{subsection}{Sect.}{Sects.}
\crefname{appendix}{Appendix}{Appendices}
\crefname{figure}{FIG.}{FIGS.}
\crefname{table}{Table}{Tables}
\crefname{equation}{Eq.}{Eqs.}
\usepackage[
  locale=US,
  separate-uncertainty=true,
  per-mode=symbol-or-fraction,
  mode = match,
  range-units = single,
  list-units = single,
]{siunitx}
\usepackage{tabularx}

\makeatletter
\def\@email#1#2{%
 \endgroup
 \patchcmd{\titleblock@produce}
  {\frontmatter@RRAPformat}
  {\frontmatter@RRAPformat{\produce@RRAP{*#1\href{mailto:#2}{#2}}}\frontmatter@RRAPformat}
  {}{}
}%
\makeatother
\begin{document}

\preprint{AIP/123-QED}

\title[A Simulation Platform for Small Solar System Bodies' Gravity Using the Einstein-Elevator]{A Simulation Platform for Small Solar System Bodies’ Gravity Using the Einstein-Elevator}
\author{E. Tahtali}
 \email{emre.tahtali@ita.uni-hannover.de}
\affiliation{Leibniz University Hannover, Institute of Transport and Automation Technology, An der Universitaet 2, 30823 Garbsen, Germany}
\author{C. Kreuzig}%
\affiliation{ 
Technische Universität Braunschweig, Institute of Geophysics and Extraterrestrial Physics, Mendelssohnstr. 3, 38106 Braunschweig, Germany
}%
\author{G. Meier}%
\affiliation{ 
Technische Universität Braunschweig, Institute of Geophysics and Extraterrestrial Physics, Mendelssohnstr. 3, 38106 Braunschweig, Germany
}%

\author{J. Blum}%
\affiliation{ 
Technische Universität Braunschweig, Institute of Geophysics and Extraterrestrial Physics, Mendelssohnstr. 3, 38106 Braunschweig, Germany
}%

\author{L. Overmeyer}
\affiliation{%
Leibniz University Hannover, Institute of Transport and Automation Technology, An der Universitaet 2, 30823 Garbsen, Germany
}%

\author{C. Lotz}
\affiliation{%
Leibniz University Hannover, Institute of Transport and Automation Technology, An der Universitaet 2, 30823 Garbsen, Germany
}%
\date{\today}

\begin{abstract}
Small solar system bodies (SSSB), which in this study are defined primarily as asteroids and comets, are becoming increasingly important as more data have become available for their study. Their significance is also highlighted by missions like NASA's OSIRIS-REx or ESA's Rosetta. However, the study of the characteristics and behavior of these objects on Earth is a challenge as the simulation of their environmental conditions are difficult. We present in this paper an approach to enable the gravity simulation of SSSBs such as comets or asteroids in a drop tower facility on Earth, which is being addressed as part of the AKUS ("Activity of Comets under Partial Gravity") project. This especially concerns gravity levels between 10$^{-2}$~\textit{g} and 10$^{-4}$~\textit{g}, where the duration of the adjusted acceleration ranges from 2.5 to 3.2 seconds. In order to simulate the conditions of SSSB as accurately as possible, an acceleration system based on servo motors and spindle axes has been developed. The accelerations are transferred from the motors to the spindle axes containing a comet-like sample. The current dimensions of the total load (including sample, sample holder, data- and communication box) are 315~mm x 160~mm x 331~mm (w x d x h), with a total weight of approximately 15~kg. These are together placed inside a vacuum chamber providing a vacuum quality of 10$^{-6}$~mbar. The whole setup is installed inside the Einstein-Elevator. Our results show that, with the current setup, we are able to generate conditions from 10$^{-2}$~\textit{g} down to 10$^{-3}$~\textit{g}. The maximum deviations under these conditions are $\pm5$$\cdot 10^{-4}$~\textit{g}. At 10$^{-2}$~\textit{g}, the duration of the experiment is at least 2.5~s limited by the travel distance of the used spindle axes, whereas at 10$^{-4}$~\textit{g} a minimum duration of 3.5~s is planned. Moreover, the experiments can be conducted at vacuum conditions of 10$^{-6}$ mbar. The results in this paper serve as a proof of concept for the generation and control of adjustable gravity levels for future SSSB experiments.  
\end{abstract}

\maketitle

\section{Introduction}\label{sec:Intro}
Space research has always played an important role for humankind. Beginning with the understanding of the night sky and the movement of stars and planets in the antiquity, the interest increased with the space age. In this context, small solar system bodies (SSSBs), defined in this study as comets and asteroids, represent a growing field of interest \cite{Hestroffer.2019}. These objects play a major role in the understanding of the solar system's formation as ancient celestial bodies, since they preserved crucial information about the origin and evolution of materials in the early solar nebula \cite{Hestroffer.2019, Pfalzner.2015, BockeleeMorvan.2017}. Additionally, they serve as essential keys to understanding the physical properties of planetesimals during the planetary accretion process.\cite{Blum.2017, Pfalzner.2015, pik.1973, Watanabe.2017} Moreover, they are relevant as future resources and a potential impact threat to Earth \cite{Britt.2023, Sonter.1997, Badescu.2013, SDRoss.2001}. Their importance is underscored by NASA and ESA missions such as "Double Asteroid Redirection Test" (DART), "Origins, Spectral Interpretation, Resource Identification, Security, Regolith Explorer" (OSIRIS-REx) or the Rosetta-mission \cite{Cheng.2023, Lauretta.2017, March.2021, Daly.2023, Schwehm.1999}.\newline
Although these kinds of missions and the gained data are very important and valuable, they also have major disadvantages. Firstly, the mentioned missions are expensive and planning-intensive. The costs can amount to several billion dollars with a planning time of a decade or more. Despite the cost and the planning, they have a high residual risk of failing. As a result, complementary methods have to be developed to make SSSB research more accessible and which support the interpretation of data gained by missions and observations. \newline 
There are some research facilities which simulate the temperature, vacuum conditions and other conditions of SSSB \cite{Kreuzig.2021,SEARS.1999,Seidensticker.1995,Bogdan.2020,Holsapple.2002,Kaasalainen.2002, J.Helbert.2024, Thomas.2017, Prince.2022}. These offer the advantage of controlled experimental environments and, therefore, the repeated study of phenomena under controlled conditions enabling important parameter studies. However, an environment of ambient \textit{g}-levels between 10$^{-2}$ and 10$^{-5}$ can rarely be provided due to Earth's overwhelming gravity \cite{Christodoulou.2024}. The gravities of asteroids for instance, such as Vesta and Psyche, are estimated approximately at 10$^{-2}$ \textit{g} \cite{Christodoulou.2024,Shepard.2017,Thomas.1997}. 67P/Churyumov–Gerasimenko's gravity, as an example comet, is estimated around 10$^{-4}$ to 10$^{-5}$ \textit{g} \cite{Basilevsky.2016,Ksanfomality.2017}. Nevertheless, it can be assumed that larger comets with higher masses, for instance C/1995 O1 or 1P/Halley, have higher gravities.\newline
Among Earth-based options, microgravity provides the closest approximation to the gravitational environment of SSSB. In this context, there are several alternative options, for instance parabolic flights, sounding rockets or drop towers\cite{Lotz.2017,Zhang.2023}. The most promising  of these are drop towers considering their microgravity quality, availability, repeatability and the flight cost. For the implementation of an additional artificial force and therefore a simulated gravity, a centrifuge or propulsion system can be used during the microgravity time of the drop tower. The Einstein-Elevator (EE), as a third generation drop tower, offers the possibility to implement of such an additional propulsion system. In contrast to drop towers of the first (pure vacuum tubes/shafts) and second generation (vacuum tube/shaft with catapult system), a small vacuum chamber, the gondola, is directly actuated by linear motors\cite{LauraFutterer.2025, Raudonis.2023}. Consequently, this design allows a high repeatability rate up to 100 flights per day and achieves a microgravity quality of 10$^{-6}$~\textit{g} (based on the evaluation methods of McPherson et al. \cite{McPherson.2017}) \cite{Lotz.2022}. Together with an implementation as formerly mentioned, an alternative opportunity is given to conduct SSSB-relevant experiments. This approach could enable high-repetition experiments in a simulated gravity environment. \newline 
In connection with this, we present a propulsion system with two drive trains each consisting of a servo motor and a spindle axis which are coupled by a sample holder. It is designed as part of the AKUS experiment (German: "Aktivität von Kometen unter partieller Schwerkraft"- English: "Activity of Comets under Partial Gravity"). The experiment involves observing the activity of comet-like samples under a vacuum of 10$^{-6}$ mbar and at adjusted \textit{g}-levels between 10$^{-2}$ and 10$^{-4}$~\textit{g} in the EE. \newline

\section{Scientific Background and Motivation}\label{sec:SDT}
SSSB research relies heavily on remote observations from Earth or data on space missions such as Rosetta or OSIRIS-REx \cite{Schwehm.1999, Lauretta.2024, Ivanova.2020, Lowry.2005}. These observations provide the foundational data necessary for further investigations, including the development of theoretical models, numerical simulations, and laboratory experiments. For instance, in comet investigations, the data collected by Rosetta are of high value. They provided insights about the physical properties, chemical composition, and the activity of the comet \cite{GabrieleE.Arnold.2017,TaylorM.G.G.T..2017, Jones.2017, Thomas.2019}. Nevertheless, there are many aspects that still have to be examined. In particular, regarding comet surface activity, several influences still have to be quantified to fully understand the ejection process of the volatiles/particles \cite{Holsapple.2002,Holsapple.2007,Thomas.2019,Bischoff.2019}. These also include the temperature and its influence caused by solar radiation, as well as the effect of the comet's gravity on the ejection process \cite{ Bischoff.2019, Jewitt.2021}. Other research aspects are the thermal properties, such as the surface and subsurface temperature distributions, which are used for modeling the thermal behavior and to gain further insights considering its activity \cite{Kreuzig.2021}. The investigation of mechanical properties of comets is essential for the physical behavior of its nuclei. For instance, the specific surface energy, or the particle radius are derived from the rolling friction force \cite{Groussin.2019}. \newline
With respect to asteroids, several physical/mechanical aspects and surface geophysics are still under investigation. Considering the first mentioned aspect, the porosity and the resulting shock dissipation are subjects of research \cite{Brisset.2018, Housen.2003}. Furthermore, regolith dynamics in milligravity is an examination subject, so that the influence of vibrations and mechanical penetration can be addressed \cite{Schwartz.2019, Brisset.2018}. Optical properties such as the bidirectional reflectance require further studies, as it is a relevant parameter for interpreting spacecraft observations \cite{J.Helbert.2024}. Also the calculation of the hemispheric emissivity for specific minerals and temperature intervals, needs to be investigated for thermophysical models\cite{Maturilli.2016}. \newline
In this context, several SSSB laboratories and SSSB-relevant platforms exist that can be used for this type of research. We define "laboratories"  as specialized facilities primarily used for SSSB research, while platforms provide certain environmental aspects relevant to SSSB. However, the latter mentioned can also be used for other types of experiments. Those laboratories and platforms allow for example the simulation of the gravity levels, the vacuum, the temperature or a combination of these. As a result, various types of experiments can be conducted under controlled environmental conditions. Based on this, important parameter studies can be done leading to improved theoretical models and calculations. In general, these laboratories and platforms can be divided into those that are subject to Earth's gravity and those where gravity can be adjusted. \newline
In the following, four examples of SSSB laboratories/platforms will be highlighted. These examples focus on different relevant environmental conditions and scopes and are connected to the work presented in this paper. One of them is the Comet Physics Laboratory (CoPhyLab) at the Technische Universität Braunschweig. Future experiments conducted on the AKUS platform will be based on those of the CoPhyLab (in \cref{subsec:CoPhyLab}). Another facility is the Planetary Spectroscopy Laboratory (PSL) in Berlin, which enables the conduction of SSSB-relevant spectral characterization experiments. In addition, the cold gas propulsion system at Beijing Drop Tower is introduced, in which SSSB research is possible. This platform provides gravity levels between 10$^{-3}$ to 10$^{-5}$~\textit{g} (in \cref{subsec:Beijing}). The controlled partial gravity platform at ZARM, in which SSSB-relevant experiments (asteroid research) have been conducted, (in \cref{subsec:Zarm}) will be presented lastly. The last two stand out since they offer adjustable gravity levels relevant to SSSB research. However, those have different acceleration concepts, sizes and experimental environments. Finally, in addition to the platforms and laboratories presented here, further laboratories and their key features are summarized in \cref{tab:labs_overview}. The chosen laboratories and platforms enable controlled experimental investigations of physical processes relevant to comets, asteroids, and other SSSBs. Moreover, their key features and research objectives are also included in \cref{tab:labs_overview}.

\subsection{CoPhyLab}\label{subsec:CoPhyLab}
The CoPhyLab is an international research project that started in 2018. The main laboratory is located at the Technische Universität Braunschweig and the project partners are the IWF Graz and the Universität Bern \cite{Kreuzig.2021, Kreuzig.2023}. The aim of CoPhyLab is to understand the basic physics behind cometary activity. Therefore, laboratory experiments with simple sample compositions are performed and supplemented by thermophysical models. In this context, micrometer-sized water-ice particles, which are the most common volatile material on comets, are produced using a custom ice machine. These particles are then studied under high vacuum conditions to observe their ejection behavior when illuminated. The heart of the laboratory is the CoPhyLab L-Chamber \cite{Kreuzig.2021}. In this chamber samples can be kept at temperatures below $150$~K and in a high vacuum of approximately 10$^{-6}$~mbar, providing the same conditions that comets experience on their approach to the Sun \cite{Groussin.2007, Tosi.2019}. The Sun is simulated using a halogen lamp. In the L-Chamber, a total of 14 instruments can be used to measure different properties of the sample and its evolution under illumination. In addition to the L-Chamber, several smaller chambers are available in the CoPhyLab-consortium, which provide the same conditions, but only have a few instruments attached. This reduces the operational complexity, allowing for parameter studies with many different kinds of samples and illumination patterns.\cite{Kreuzig.2025} \newline

\subsection{Planetary Spectroscopy Laboratory}\label{subsec:PSL}
For a spectral characterization of diverse analog materials for asteroids, comets, and icy moons, the Planetary Spectroscopy Laboratory (PSL) can be considered. The facility conducts bi-directional reflectance and emissivity measurements within a high-vacuum, cryogenic environment. This provides the essential "ground truth" needed to identify the mineralogy and chemical composition of these bodies from remote sensing mission data. It is located at the German Aerospace Center (DLR) in Berlin. The results of the measurements will be applied to future missions such as ESA JUICE or NASA Europa Clipper. \newline
The laboratory consists of a compact cryogenic simulation chamber designed for angle-dependent-directional reflectance measurements. The chamber allows up to four samples simultaneously to be kept. These are investigated at temperatures down to approximately 63~K in a vacuum environment of down to 10$^{-6}$~mbar. The laboratory provides conditions representative of SSSB and shadowed regions of the Moon. With the installed spectrometer (Bruker Vertex 80V), measurements can be performed with a wide spectral range from ultraviolet to far-infrared. Moreover, the setup integrates a nitrogen-purged glovebox and an airlock system, so that icy samples or ice-dust mixtures can be prepared at temperatures between 223~K and 183~K. Consequently, these features prevent degradation of the vacuum quality and minimize the development of frost contamination inside the chamber. As an external light source, the PSL uses halogen lamps to simulate solar radiation.\cite{J.Helbert.2024}   

\subsection{Cold gas propulsion system at Beijing Drop Tower}\label{subsec:Beijing}
The necessity of adjustable gravity ranges for SSSB research has been indicated in the introduction of \cref{sec:SDT}, especially regarding their surface activity \cite{Bischoff.2019, Jewitt.2021, Schwartz.2019, Brisset.2018}. A propulsion system that is capable of generating gravity levels from 10$^{-3}$ to 10$^{-5}$~\textit{g} has been developed at the Beijing Drop Tower, consisting of an inner and outer capsule (two-capsule system)\cite{Zhang.2023, Liu.2016, Wan.2010}. During the 3.6~s long drop of the capsule, cold-gas thrusters are activated, which are attached to the bottom of the inner capsule of the system. The generated acceleration is measured using an acceleration sensor, which is placed inside the inner capsule. The accelerometer has a measurement range of $\pm2$~\textit{g}, a resolution of $1\cdot10^{-6}$~\textit{g}, and a bias of less than 3~m\textit{g}. Additionally, it has a sensitivity of $1.2 \pm 0.2\ \mathrm{mA}/g$ and a natural frequency of $800\ \mathrm{Hz}$. \cite{Zhang.2023} \newline
The thrusters, which use de Laval nozzles, vary in their geometries to generate the desired accelerations. In total, 24 thrusters have been installed with different amounts of thrust, specifically four \SI{50}{mN} and 20 \SI{100}{mN} thrusters, so that the applicable thrust ranges between \SI{200}{mN} and \SI{2,000}{mN}. To achieve 10$^{-4}$ \textit{g}, the group of 50 mN thrusters is activated, while all \SI{100}{mN} thrusters are used to achieve a gravity level of 10$^{-3}$~\textit{g}. \cite{Zhang.2023}
Experimental results show that, during free fall, the residual acceleration of the inner capsule reaches approximately $6\cdot10^{-5}$~\textit{g} in the falling and about $1\cdot10^{-5}$~\textit{g} in the transverse directions. For the experiments regarding an adjusted gravity of 10$^{-4}$~\textit{g}, the acceleration data show, that the acceleration starts at 10$^{-3}$~\textit{g} and decreases hyperbolically to approximately $2\cdot10^{-4}$~\textit{g} after 2~s. The subsequent progression remains nearly constant for an additional 1.5~s. With respect to 10$^{-3}$~\textit{g}, the achieved acceleration is approximately within the targeted acceleration at the begin but shows a slight decreasing trend from 0.5~s to 2.5~s. After that, the acceleration decreases more significantly to the 10$^{-4}$~\textit{g}-regime. \cite{Zhang.2023} Short-duration acceleration disturbances are observed, for instance at thruster activation and shut off. Moreover, spectral analysis identified a dominant micro-vibration at 43 Hz caused by the capsule due to the release process. \cite{Zhang.2023}\newline 
However, results of a test flight, in which all thrusters have been activated, have not been published yet. Moreover, due to the design of the nozzle layout, it is difficult to adjust different \textit{g}-levels by activating the nozzles separately. This is because of a rotational force that could act on the experimental platform and could harm the infrastructure. Although the platform, as described, can enable gravity levels of SSSB, research or experiments in this area have also not been published yet.\cite{Zhang.2023}

\begin{table*}[] \centering \caption{Some SSSB-relevant laboratories and platforms, their location, main features/characteristics, and investigated properties.} \label{tab:labs_overview} \begin{tabularx} {\textwidth}{p{2cm} p{1.5cm} p{3cm} X X } 
\hline Facility & Location & Target objects/application & Features/characteristics & Investigated Properties \\ 
\hline CoPhyLab L-Chamber\cite{Kreuzig.2021} & TU Braunschweig (Germany) & Comets & Cryogenic vacuum chamber ($10^{-6}$\,mbar) with temperatures of 150~K, solar illumination, and further instruments (several different cameras, temperature sensors, mass spectrometer,...) & Investigations regarding the cometary activity, Particle ejection, sample evolution, thermal transport \\ 
\hline Planetary Spectroscopy Laboratory (PSL)\cite{J.Helbert.2024} & DLR Berlin \newline (Germany) & Asteroids, comets and Icy moons & Cryogenic vacuum chamber, light source with spectral range covering UV to FIR/TIR & Spectral measurements especially regarding the bi-directional reflectance \\ 
\hline Cold gas propulsion system \cite{Zhang.2023} & Beijing \newline (China) & Comets and Asteroids possible; however, no SSSB-specific experiments reported & Adjustable gravity levels from $10^{-3}$--$10^{-5}$\,\textit{g} with a cold gas propulsion system used at the Beijing Drop Tower & Potential for SSSB experiments under adjustable gravity \\ 
\hline Controlled Partial Gravity Platform\cite{Joeris.2025} & ZARM Bremen \newline (Germany) & Asteroids & Adjustable gravity levels from $10^{-2}$ to $10^{-4}$\,\textit{g} with a linear drive system with the Bremen drop towers & Impacts, granular dynamics \\ 
\hline IMPACT Laboratory\cite{Thomas.2017} & University of Colorado Boulder \newline (USA) & Micrometeoroids / Interplanetary dust particles (comet/asteroid fragments) & 3-MV electrostatic dust accelerator in a vacuum environment & Micrometeoroid ablation and ionization coefficients ($\beta$) for various gases \\ 
\hline Center for Materials Interfaces in Research and Applications (MIRA)\cite{Prince.2022} & Northern Arizona University \newline (USA) & Asteroids & Nd-YAG pulsed laser (1064 nm), ultra-high vacuum chamber ($10^{-8}$\,mbar) & Space weathering of hydrated minerals (pentlandite, phyllosilicates), thermal-vacuum testing, solar simulation \\ 
\hline Physics Laboratory Warsaw\cite{Kossacki.2022} & University of Warsaw \newline (Poland) & Comets & Vacuum chamber ($10^{-6}$\,mbar), active cooling to approximately 150~K and a specialized infrared (IR) radiator, laser distance monitoring for observing purposes & Outgassing, surface recession, and dust mobility \\ 
\hline ICE Setup \cite{Brisset.2022} & University of Central Florida \newline (USA) & Icy moons, bodies of Saturn's rings, no asteroid or comet research reported & Vacuum chamber with cryogenic cooling (130--140~K), drop mechanism for projectiles ($< 4$~m/s), laser sheet illumination for ejecta tracking by a high-speed camera & Low-velocity impacts into icy regolith, crater scaling laws, ejecta mass production \\ 
\hline ISAE-SUPAERO Drop Tower\cite{Sunday.2016} & Toulouse \newline (France) & Asteroids (landing and sampling missions), but also comets & 5.5~m drop tower frame using an Atwood machine (pulleys/counterweights) to create adjustable gravities of $10^{-1}$ to $10^{-2}$\,\textit{g} for approximately 1~s & Low-velocity collisions (up to 20~cm/s), regolith strength and mechanical response \\ 
\hline TEMPus VoLA\cite{Capelo.2022} & University of Bern \newline (Switzerland) & Comets, protoplanetary disks & Multi-pressure vessel for microgravity platforms (parabolic flights), dust injector, high-speed shadowgraphy & Collective dust-gas aerodynamics, gas permeability of high-porosity media, Epstein drag coefficients of dust aggregates with different velocities\\ \hline SCITEAS (LOSSy Lab)\cite{Pommerol.2015} & University of Bern \newline (Switzerland) & Comets, Mars and icy satellites & Thermal vacuum chamber ($10^{-7}$~mbar), Temperature ranges: 165 to 205 K, cooling mechanism based on a cold shroud, solar simulator (Xenon lamp), VIS/NIR hyperspectral imaging & Spectro-photometric and morphologic evolution of sublimating icy analogues, sintering, dust mantle formation \\ 

\hline 
\end{tabularx} 
\end{table*}

\subsection{Controlled partial gravity platform at ZARM}\label{subsec:Zarm}
Another platform, in which SSSB-relevant experiments have already been conducted, has been developed for the GraviTower (GTB) and the ZARM drop tower in Bremen, namely the "Controlled Partial Gravity Platform for Milligravity" \cite{Joeris.2025, Kampen.2006, Konemann.2022}. The platform is built for asteroid impact investigations. During microgravity conditions provided by the drop towers, it generates an additional artificial force using a linear motor. The system consists of several components beyond the actuation components, such as a vacuum chamber with the necessary pumping stages, a launcher that pushes the impactor onto the granular bed, and cameras together with a lighting system that observe the impact. By launching particles into a regolith bed under milligravity, one aim is to examine granular interactions in the cohesion-dominated regime. \newline
Its acceleration system consists of one linear motor translation stage, which is typically used for applications in the field of optics. During the microgravity time, the linear stage accelerates the vacuum chamber together with the launcher. The travel distance of the system is 300 mm, enabling different experiment times depending on the adjusted acceleration and the drop tower used \cite{Joeris.2025}.  The limits resulting from its length are primarily found in the higher acceleration ranges. If, for instance, an acceleration of 10$^{-2}$~\textit{g} is desired, the experiment time is below 2~s.\newline
As initially mentioned, the platform can be used at the two drop towers in Bremen, whose main differences are the repetition rate and microgravity duration. The GTB provides a repetition rate of several tens of launches per day and has a microgravity duration of 2.5~s. In contrast, the ZARM drop tower offers a microgravity time of 9.3~s. However, it only allows two experiments per day. The available microgravity duration is also the limiting factor concerning the use of the partial gravity platform. At adjusted gravity levels in the lower regime, (around $10^{-4}~\textit{g}$) an experimental time of the 9.3~s is possible, as the resulting travel distance is smaller, too. Consequently, the travel distance of 300 mm is not a limiting factor at those gravity levels. As a result, experimenters can choose between a shorter microgravity time with high repetition, or a longer microgravity time with fewer repetitions. \newline
The acceleration is controlled via the encoder of the motors. According to the published results, the linear stage follows an adjusted trajectory based on the chosen gravity level. The data show small sub-millimeter deviations from the ideal parabolic trajectory, as inferred from encoder-based position measurements \cite{Joeris.2025}. However, additional sensor data, such as accelerometer data have not been published yet, even though they directly indicate the prevailing acceleration in the experiment area. This also applies to accelerations based on the encoder data. 

\begin{figure*}[t!]
	\includegraphics[width=15cm]{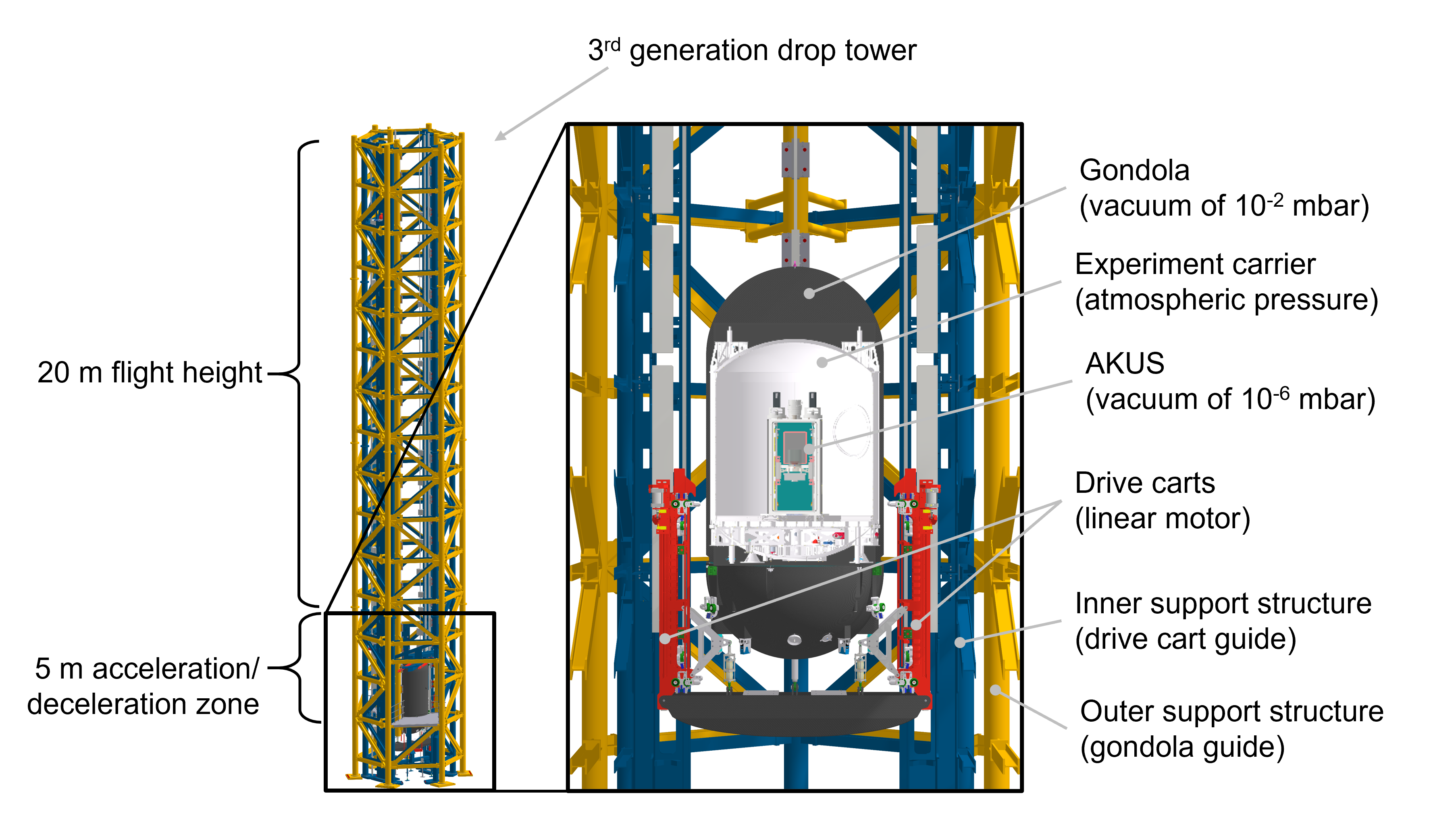}
	\caption{The Einstein-Elevator (EE) and its main components.}
	\label{fig:EE}
\end{figure*}

\section{Experimental Platform and Setup}\label{sec:AKUS}

To conduct the planned experiments within AKUS, a microgravity environment is necessary. This allows for the application of an artificial propulsion system which generates the relevant accelerations similar to \cref{subsec:Zarm} and indicated in \cref{sec:Intro}. The platform which provides this condition is the EE, a third generation drop tower. Therefore, the EE as the experimental environment of AKUS will be described first before the actual experimental setup of AKUS.

\subsection{The Einstein-Elevator}
The EE is a novel drop tower, which is used for quantum sensing, space and space-production related research\cite{Anton.2025, Garcion.2025,Raffel.2024, Raupert.2021}. It is located at Leibniz University Hannover. Due to its unique design, a microgravity quality of 10$^{-6}$~\textit{g} as well as other gravity levels such as those on the Mars or Moon surface are achievable\cite{Lotz.2022}.
The facility is able to perform 300 flights per day, with 100 flights per 8-h shift. The high repetition rate is enabled by the system design, in which only the gondola is evacuated to achieve vacuum. Consequently, acoustic decoupling is achieved significantly faster than earlier-generation drop towers, where the entire drop tube has to be evacuated. Additionally, the EE is able to accelerate a payload with a total mass up to 1,000~kg and has a high degree of automation. Furthermore, it is able to contain experiments with a height of 2~m and a diameter of 1.7~m.\ Those key aspects are enabled by three main components that play a crucial role: the drives, the guidance system and the gondola that includes the experiment carrier within a vacuum presented in \cref{fig:EE}.\newline
For the simulation of microgravity, the flight begins with a 5~\textit{g} acceleration phase of 0.5~s, which is provided by the drives. A short deceleration phase follows to create a distance between the carrier and the gondola. This is the starting point for the microgravity period of 4~s. With the end of the microgravity time, the distance between the carrier and the gondola is reduced to zero again, so that the carrier and the gondola are reconnected. Afterwards, the gondola is decelerated with 5~\textit{g} for 0.5~s, so that the gondola returns to its launch position \cite{Lotz.2017}. Finally, the experiment carrier is centered again due to displacements caused by the Coriolis forces during the free-fall time and ending the flight.\newline
For our experiment, a high microgravity quality is necessary, as this is fundamental to adding an artificial force, which simulates the gravity conditions of SSSB. The EE provides a high microgravity quality of 10$^{-6}$~\textit{g}, as the gondola and the drive guidance are each mounted on separate towers in a tower-in-tower design. This separation reduces the transmission of vibrations into the experiment carrier, mainly caused by the drives. The motors and the gondola are connected only by a coupling rod. In addition, the foundations of both towers are separated from each other. The experiment carrier, designed as a pressure vessel, is important because it enables vacuum outside itself and thus inside the gondola\cite{Sperling.2023}. As a consequence, experiments are at atmospheric pressure inside the carrier with an acoustic decoupling given by the vacuum outside of it, reducing the transmission of vibrations further. \newline
Moreover, the drop tower is able to generate gravity conditions from 0.1~\textit{g} to 1~\textit{g} (hypogravity) and 1~\textit{g} to 5~\textit{g} (hypergravity). In contrast to microgravity flights, the experiment carrier has to be attached to the gondola ground for these gravity levels. An adjustment of the trajectory of the gondola is required for the desired acceleration from 0.1~\textit{g} to 1~\textit{g}, directly influencing the duration of the test time. However, gravity levels of SSSB cannot be simulated by the EE's drives, as vibrations inside the experiment carrier, which is in this case fixed to the gondola floor, are exceeding in amplitude the constant low accelerations of interest. Those vibrations are caused by the motors, the roller guides and other components of the Einstein-Elevator and are directly transmitted into the experiment. Therefore, a decoupling of the experiment from the gondola floor and the additional axis-based drive system by-passe this issue. \newline
Additionally, experiments can be conducted from the top starting position of the tower as an alternative option. Therefore, initial disturbances of the 5~\textit{g}-acceleration phase can be avoided for acceleration-sensitive experiments, leading to a smoother and direct transition from 1~\textit{g} to 0~\textit{g}. Consequently, this also halves the microgravity time to 2~s.

\begin{figure*}[t!]
  \centering
  \makebox[\textwidth][c]{
    \begin{minipage}[b]{0.48\textwidth}
      \centering
      \includegraphics[height=7cm]{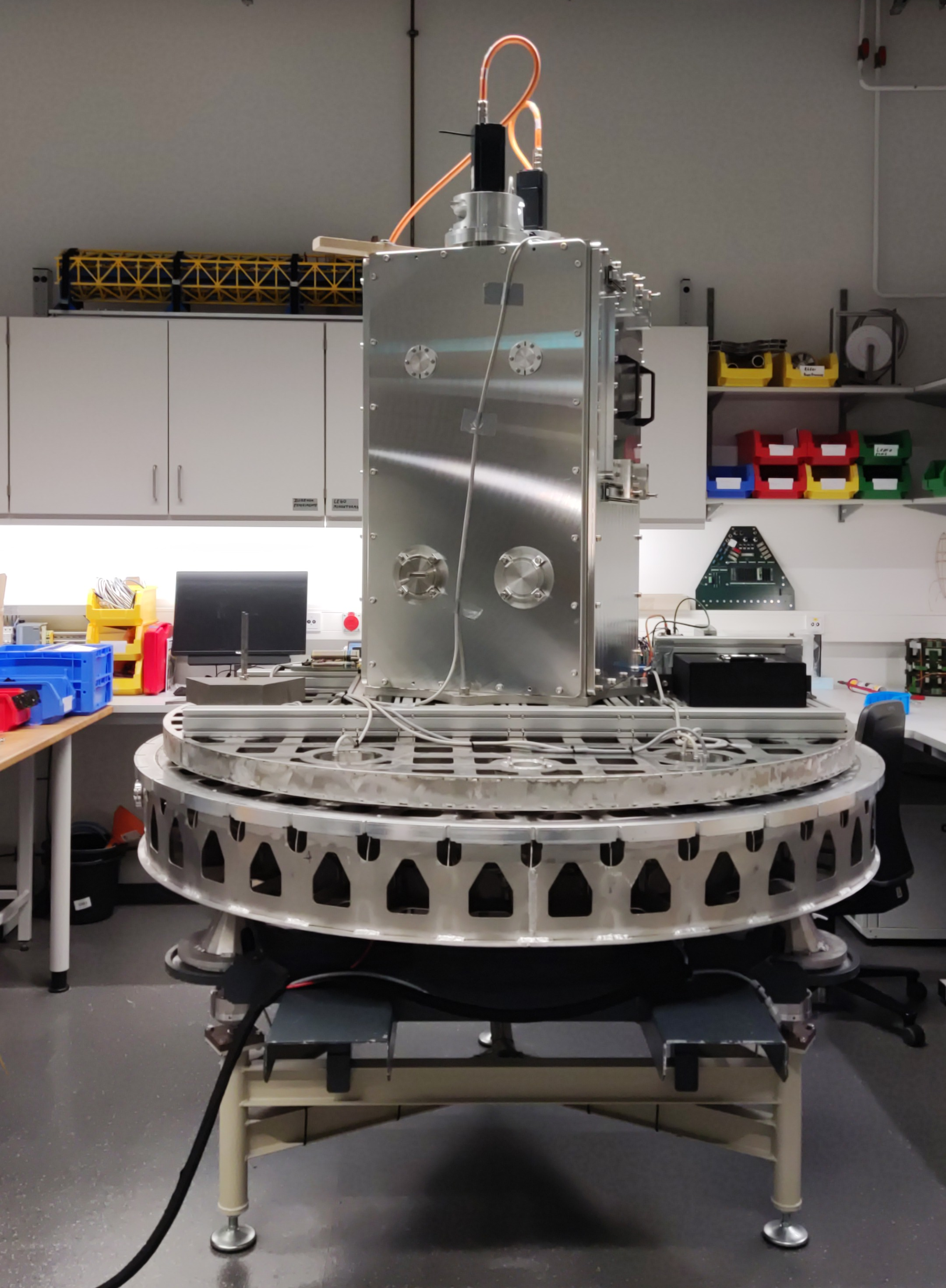}
      \\[1ex]
      (a) The AKUS system build up on the bottom part of the experiment carrier.
      \label{fig:wideA}
    \end{minipage}%
    \hspace{0.00001\textwidth}
    \begin{minipage}[b]{0.48\textwidth}
      \centering
      \includegraphics[height=7cm]{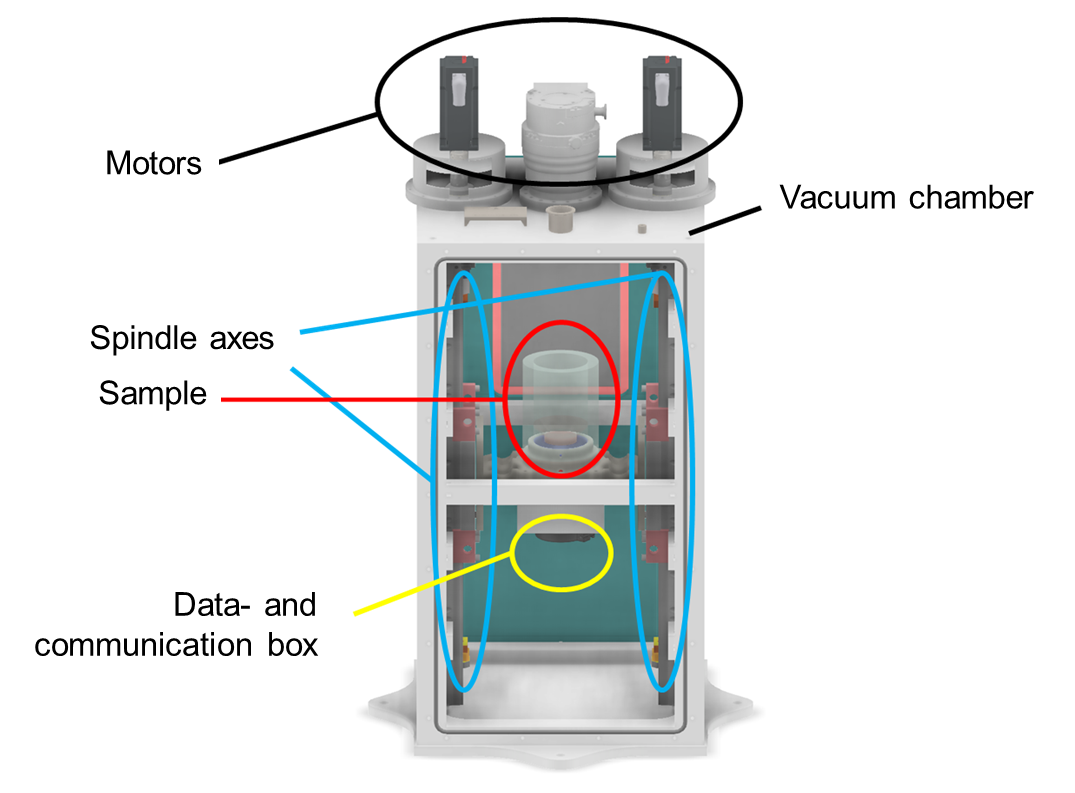}
      \\[1ex]
      (b) The main components used for AKUS.
      \label{fig:wideB}
    \end{minipage}%
  }
  \caption{The AKUS system and its components. The whole system (including the payload level) has a diameter of approximately 1700~mm and height of 1400~mm. Together with the experiment carrier, it weighs approximately 950~kg. The sample has a diameter of 140~mm and height of 150~mm. Its weight is approximately 1~kg. }
  \label{fig:AKUS_EE}
\end{figure*}

\subsection{Experimental setup of the AKUS experiment}
Within this subsection, a general description of the AKUS-experiment will be given firstly. Afterwards, the acceleration system will be described in detail, as it is necessary to provide the gravity conditions of SSSB. Finally, the current graphical user interface (GUI) will be presented, since it will play a major role in the future regarding the monitoring of the system.

\subsubsection{General description of the Experiment}
\begin{figure}[]
	\includegraphics[width=8cm]{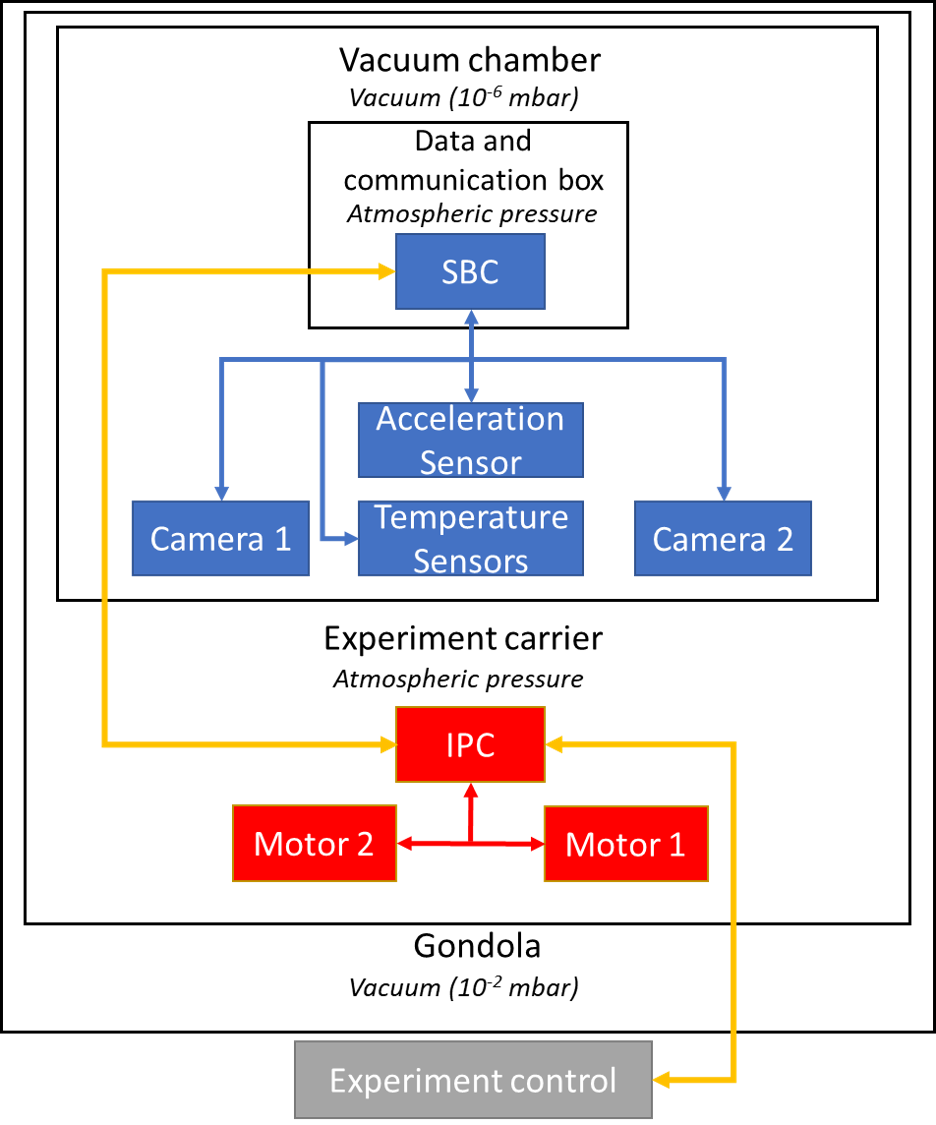}
	\caption{General overview regarding the control components and their connections to each other}
	\label{fig:G_Overview}
\end{figure}

The scientific goals of AKUS build on the experiments conducted inside the L-Chamber of the CoPhyLab, but additionally utilizing the microgravity environment of the EE. Consequently, AKUS will allow cometary particle ejection experiments under low gravity conditions in the EE. In contrast to former experiments conducted under Earth gravity, these experiments will allow the influence of gravity on the ejection process to be investigated. \newline
To simulate the environmental conditions of comet surfaces in the EE, four main aspects of the experimental setup have to be highlighted, namely the mechanical components, the actuators, the sensors and the systems for data processing and communication. The first aspect consists of the vacuum chamber, which is used to achieve vacuum conditions up to 10$^{-6}$~mbar. The vacuum chamber is mounted on the experiment carrier of the EE, which is operated at atmospheric pressure, seen in \cref{fig:EE}. Moreover, it contains several mechanical (e.g. for the actuators or the vacuum pump) or electrical interfaces (connection points for sensors and other electronic components).\newline
There are a total of two drive trains, each consisting of a motor, a rotary feedthrough and a spindle. These two drive trains are connected to each other with a sample holder made of titanium, which was generatively designed for reasons of stability and low weight. Due to symmetrical design, rotational influences during the microgravity time are minimized. The motors are placed outside of the vacuum chamber and connected via the feedthroughs to the vacuum-compatible spindles. The entire system, including the payload level, measures approximately 1700 mm in diameter and 1400 mm in height. Combined with the experiment carrier, its total weight is around 950 kg. The total load, attached to the spindle carts, is approximately 15 kg and its dimensions are 315~mm x 160~mm x 331~mm (w x d x h) including sample,  sample holder, and the data- and communication box, seen in \cref{fig:AKUS_EE}. During the free fall phase (approximately 0.5 s after the 5~\textit{g} acceleration phase), the comet-like samples will be accelerated with the desired gravity level of SSSB, which is in the range of 10$^{-2}$ to 10$^{-4}$~\textit{g}. 
\newline While this acceleration is applied, the sample behavior will be observed by a camera (Daheng-Imaging; MER2-160-75GC), which generates up to 75~fps. 
In addition to the cameras, other sensors are also installed such as temperature sensors connected to the sample, and an acceleration sensor mounted on the sample holder(\cref{fig:G_Overview}). All sensors are placed outside the data and communication box in the vacuum environment. The differential accelerometer (Kistler 8396A) has a measurement range of $\pm 2~\textit{g}$ and a threshold resolution of $1\cdot10^{-4}$~\textit{g} (more specifications in \cref{tab:accelerometer_specs_AKUS}).  \newline 
Moreover, the servo motors have encoders to control motor motion. With those, the motor positions are controlled as a function of time to achieve the desired accelerations. In the long term, the controlled variable will be the acceleration signal measured by the accelerometer as its resolution is sufficient ($1\cdot10^{-4}~\textit{g}$). Nevertheless, the effectiveness of such control strongly depends on the design of the actuator control loop. 
\newline 
For acquiring data from the different sensors, a single-board computer (SBC) will be installed  inside the data- and communication box in the future(\cref{fig:G_Overview}). It is a vacuum-compatible, sealed pressure chamber, which has atmospheric pressure in the inside. This pressure chamber will be mounted directly under the sample holder in the vacuum environment and contains several other electrical components, for instance, AD or step-down converters. The data transfer from the SBC to the industrial PC (IPC), which controls the drives, is planned to be contactless. Possible data transfer methods still need to be tested and examined regarding their capabilities and transfer rate. At the moment, the data transfer is enabled with cables.

\begin{table}[htbp]
  \centering
  \caption{Key specifications of the differential accelerometer (Kistler 8396A)}
  \label{tab:accelerometer_specs_AKUS}
  \begin{tabular}{l c}
    \hline
    Parameter & Value \\
    \hline
    Measurement range & $\pm 2\,\textit{g}$ \\
    Noise density & $0.0007\,\mathrm{mg_{rms}}/\sqrt{\mathrm{Hz}}$ \,(0--100\,Hz) \\
    Threshold resolution & $0.1\,\mathrm{mg_{rms}}$ \,(=$1\times10^{-4}\,\textit{g}$) \\
    Sampling rate & $5{,}000\,\mathrm{Hz}$ \\
    \hline
  \end{tabular}
\end{table}

\subsubsection{Detailed Design of the Acceleration System }\label{subsubsec:Acceleration}

The spindle axes used for our system have a travel distance of 680 mm. The choice of the length depended on two factors: firstly, the 4~s microgravity time of the EE, in which the whole process has to be conducted. Besides the constant acceleration, this includes the deceleration time of the carts. Secondly, it depended on the maximum travel distance resulting from the maximum acceleration of 10$^{-2}$~\textit{g}, which is 600 mm at 3~s of acceleration time. Smaller accelerations are only limited by the microgravity time, as the travel length is long enough for several seconds of conducting time.\newline 
In general, servo motors are not designed to achieve constant acceleration values, but rather to maintain a defined rotational speed or to achieve precise positioning. However, in our case, a stable acceleration torque is required to produce a constant linear acceleration. In the following, we will therefore briefly review the motion characteristics of servo motors and the relevant mechanical optimizations of our acceleration system.

As described before, our servo motors are actuating the spindle axes and for the calculation of the desired torque, the following equation is the standard: 

\begin{equation}
 M_T= M_L+ M_A.
\end{equation}

Here ${M_T}$ is the total torque, that the motor has to supply. ${M_L}$ is the resulting torque caused by the cart load, friction and resistance of the drive train components. It is mostly static and has no major influence on the acceleration of the system. In our case, ${M_L}$ was calculated conservatively (with safety margins) and equaled approximately 0.5~Nm. Thus, it remains well below the motor's nominal torque of 1.3~Nm. \newline
Lastly, ${M_A}$ is the acceleration torque and the most important in our case. The importance of the acceleration torque can be seen in the following equation:
\begin{equation}
 M_A= J_T\cdot \alpha, %
\end{equation}
in which $J_T$ is the total inertia and $\alpha$ the angular acceleration of the motor. In addition, $\alpha$ can be substituted as follows:

\begin{equation}\label{eq:2.2}
 M_A= J_T\cdot\frac{\pi\cdot \Delta n}{30 \cdot\Delta t_A}. %
\end{equation}

\begin{figure}[htb]
	\includegraphics[width=8cm]{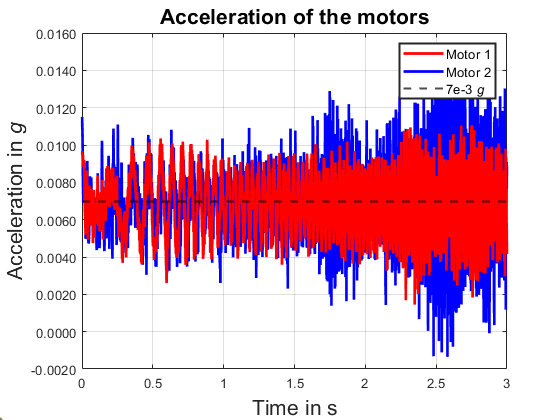}
	\includegraphics[width=8cm]{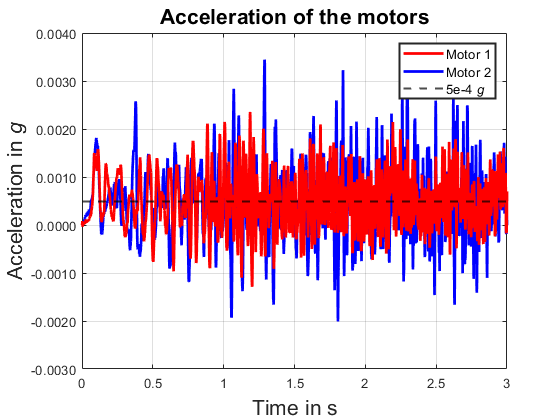}
	\caption{Behavior of the motors regarding for 7$\cdot 10^{-3}$~\textit{g} (top) and 5$\cdot10^{-4}$~\textit{g} (bottom). The data is derived from the encoders of the motors. The rotational acceleration measured by the encoders has been converted into linear acceleration. For this, the relevant parameters of the spindles, such as lead or pitch, and the rotational speed have been taken into account. As a result, an understanding of the linear motion characteristics of the spindle carriage can be derived by the rotational data.} 
	\label{fig:OS}
\end{figure}

$\Delta n$ is the difference of the rotational speed that should be achieved during the acceleration time $\Delta t_{A}$. As a consequence, it is the key factor that provides the desired acceleration of the sample. Considering our system without adjustments, the total inertia $J_T$ of the load and the motors is approximately $1.7\cdot10^{-4}$ kg$\cdot$m$^2$. Based on the spindle lead, the rotational speed \textit{n} accelerates from 0 to $\approx$ 3600~$/$min within a given time of 3 s for a desired linear acceleration of $10^{-2}~\textit{g}$. \newline 
Based on these assumptions and using the \cref{eq:2.2}, the acceleration torque results in $\approx$ \SI{0.0214}{Nm}. With regard to our servo motor, the acceleration torque corresponds to $\approx 1.65\%$ of the nominal torque (\SI{1.3}{Nm}). According to the manufacturer, the current controller can quantize to $1/1,000$ of the nominal torque. However, they suggest that for effective acceleration control, at least 100 current increments are necessary (\SI{0.13}{Nm}). As a result, the motor's inner current controller cannot reliably produce constant and small torques/accelerations as the acceleration torque corresponds to $\approx 16.5\%$ of the necessary torque for 100 current increments (\SI{0.13}{Nm}). Hence, a constant acceleration cannot be realized with sufficient resolution due to the small torque. This is also a result of the small motor inertia. 
\newline \cref{fig:OS} shows the acceleration data derived from the encoder data. In connection with this, an acceleration overshoot and rapid fluctuations around the target value are seen. Although this can be a result of control-law dynamics, the former mentioned torque resolution with a closed-loop controller and low inertia can also produce a similar behavior. The reason is the cascade control design, in which the current controller (inner cascade) and the position/velocity controller (outer cascade) are used for motor control. Even if a discretization is not seen in the encoder data, as it observes the outer loop, a limited motor current resolution could potentially also cause such behavior. Additionally, servo motors and their controllers are primarily designed for position and velocity purposes. Torque control, especially in the lower regime to generate accelerations between $10^{-2}$ and $10^{-4}$~g, is not necessary, so the current resolution does not need to be very high. 

One possibility to overcome this challenge is to increase the inertia of the system due to the framework conditions with respect to the microgravity time and the rotational speed based on the spindle axes. However, it should be emphasized that this approach also has some limitations in form of the inertia ratio $\lambda$ defined as the ratio of the motor's rotational inertia $J_M$ to the resulting inertia given by the load $J_L$ \cite{Li.2023} as seen in \cref{eq:2.3}:

\begin{equation}
 \lambda= \frac{J_L}{J_M}\label{eq:2.3}. %
\end{equation} \newline

According to various motor developers, an inertia ratio of $\lambda \leq 5$ should be targeted for a good motion behavior and control\cite{.22.08.2025, .22.08.2025b}. For this reason, the motors used here have adapted longer shafts, which already increases the motors' inertia. Due to the longer shafts, it is also possible to mount flywheels on the shafts. As a result, the motor inertia and consequently the total system inertia increase. These aluminum flywheels have a diameter of around 15~cm with a resulting inertia of $J_F \approx 1.210\cdot10^{-3}\,\si{\kilogram\metre\squared}$. Based on \cref{eq:2.2}, the resulting $M_A$ is $\approx$ \SI{0.17}{Nm}, so that at least 133 increments are given, which is, as stated before, necessary for a good control of the acceleration. Furthermore, the inertia ratio $\lambda$ can also be improved. Without consideration of the flywheels, the value of $\lambda$ was approximately 26, changing with the improvement to a value of approximately 1. The described mechanical adjustments have been used for the results presented in this manuscript in \cref{sec:results}.

\subsubsection{Graphical user interface}
\begin{figure}[htb]
	\includegraphics[width=9 cm]{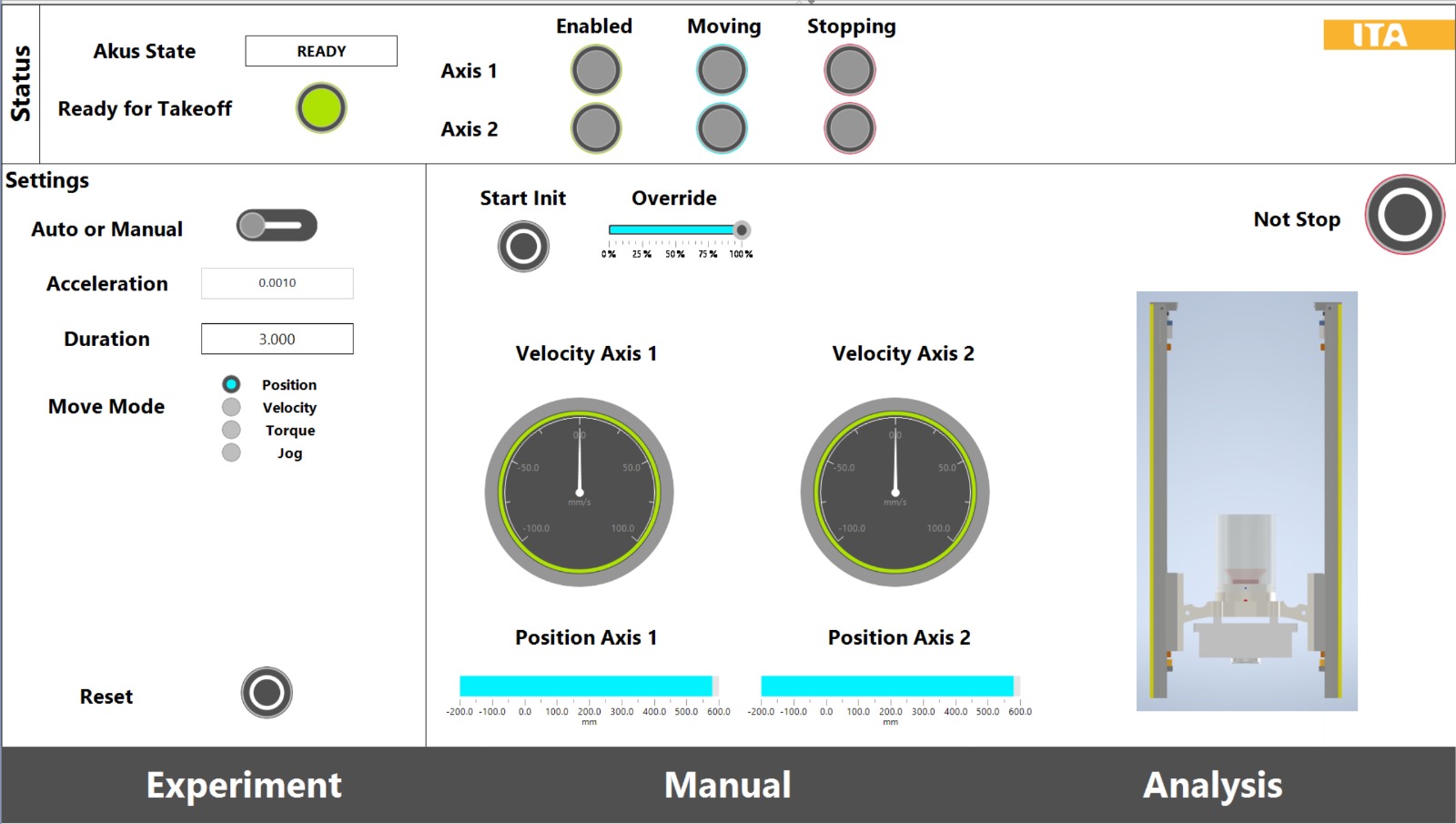}
	\caption{GUI of the experiment. Through it, the desired accelerations and also their duration can be set. Moreover, the movement of the spindle carts with the sample holder can be observed simultaneously via the digital twin.}
	\label{fig:GUI}
\end{figure}

To execute the experiment, a graphical user interface (GUI) has been developed (\cref{fig:GUI}). The GUI was built using TwinCAT, as the IPC and several automation components are Beckhoff products. The current GUI contains several features to simplify the execution of the experiments. It is divided into three sections: experiment, manual and analysis. 
\newline The experiment section includes the collection of the flight data or sets the desired acceleration with its belonging time. Moreover, a simplified digital twin that maps the movement of the spindle carts during an experiment is also implemented in the GUI; it is currently  used for monitoring purposes only.
\newline The control of the motors allows different modes which are selectable. Those are position, velocity, torque and jog. To conduct experiments, the motors have to be enabled which is indicated by the status lights in the upper part of the GUI. Then, a reference run has to be performed once with the "Start Init" button of the GUI. This determines the zero position of the spindle carts via limit switches. During motion, the corresponding "Moving"-lights will be activated and the cart positions are displayed by blue progress bars on the lower part of the GUI. The center area shows the speed indicators for the current velocities of the axes. 
\newline The manual section consists of all functions of the former section; however, implements tools for testing and adjustments additionally. The manual mode of the GUI allows for instance to perform manual positioning, as this is important for sample changes or adjustments purposes. To enable this, the manual switch has to be activated (left side of the GUI) and the jog mode must be selected. Thereafter, manual positioning can be done via jog buttons and an override slider to set the manual velocity. \newline 
The analysis section contains the presentation of the encoder and acceleration data of the current experiment. The "Not Stop"(emergency stop) button immediately halts the system to protect hardware and experimenters during tests or flights. \newline 
In the future, the GUI-functions and also the digital twin will be expanded. As a result, simulations will be possible for safety purposes or generate artificial data. Using this data gained by the simulation, machine learning techniques could be applied and trained for monitoring purposes and an improved controlled system. 

\begin{figure*}[t!]
	\includegraphics[width=19cm]{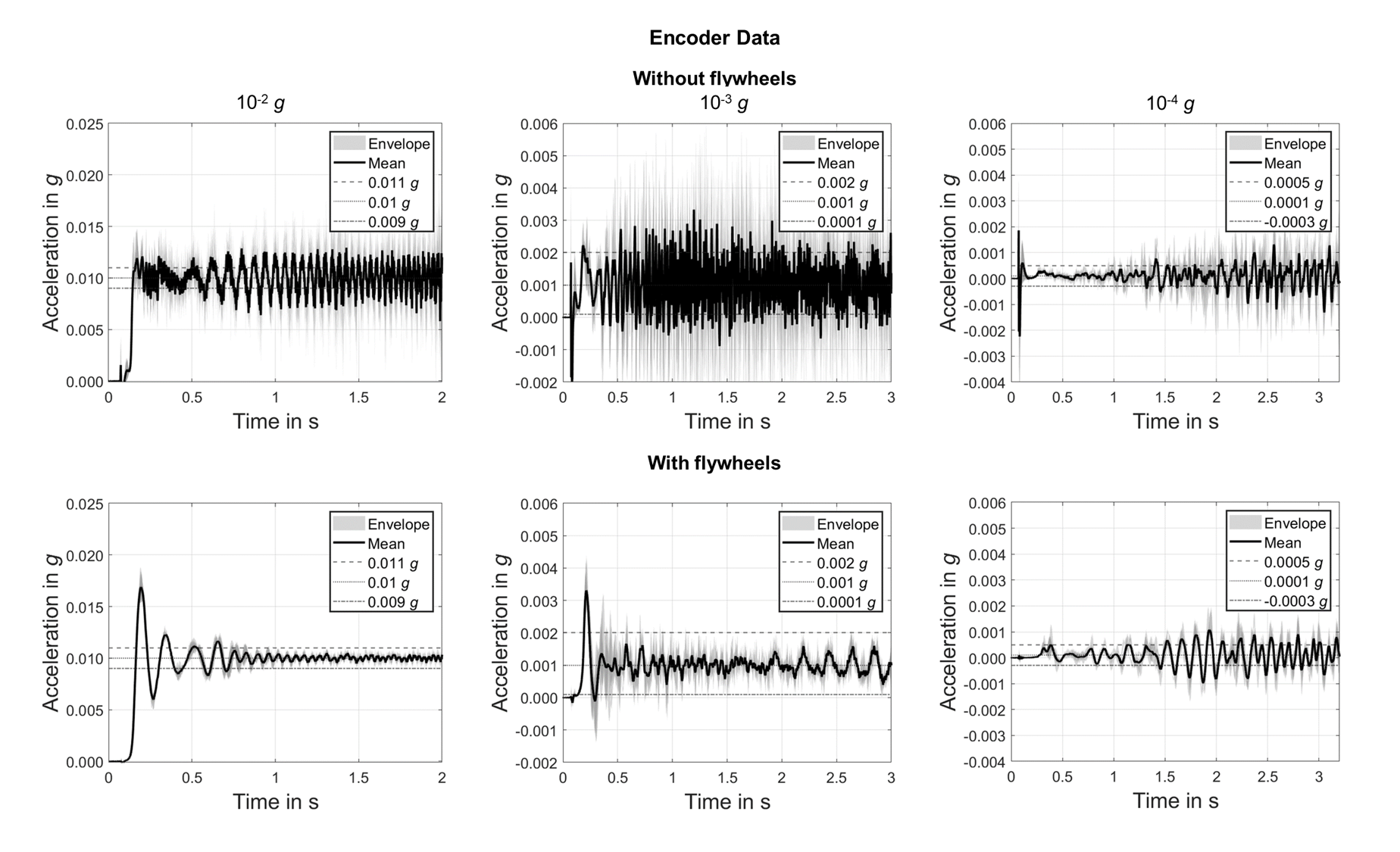}
	\caption{Comparison of the encoder data without (top) and with (bottom) flywheels. The mean consists of the encoder data from both motors and the average of five flights. The envelope is defined by the maximum/minimum values of the means from both encoders across the considered five flights. The encoder data demonstrate a positive effect of the flywheels, enabling a better control of the desired accelerations. The data was not filtered.}
	\label{fig:Flywheel}
\end{figure*}

\section{Results} \label{sec:results}
This section presents various results regarding the acceleration of the system based on the accelerometer and the encoder data. Firstly, the results under normal Earth gravity for an analysis of the influence of the flywheels compared to conditions without (in \cref{subsec:EE_flywheel}) will be presented based on the encoder data. After that, the results of experiments follow, which have been conducted under Earth's gravity and in microgravity inside the EE (in \cref{subsec:ex_micro}). For these test campaigns, the experimental setup of AKUS consisted of the main components including the vacuum chamber, the spindle axes, the motors, and the acceleration sensor. The vacuum chamber was at atmospheric pressure, as vacuum is only needed for tests with the comet-like samples. Therefore, no samples were used in these tests. Moreover, cables were not routed as planned in the final design, although these are sources of vibrations which may affect the quality of the adjusted accelerations\cite{Banerjee.1996, Fialho.04032000, Ardelean.2015}. Especially at the EE, these vibrations occur during the transition from 5~\textit{g} to $\mu\textit{g}$, during which the gravitational load on the cables are suddenly released. In addition, during the tests inside the EE, the gondola was not evacuated, resulting in the experiment carrier operating in a normal pressure environment due to limited experimental time available. Since the vacuum in the gondola is mainly used for acoustic decoupling, the normal pressure environment could induce vibrations from any components of the EE inside the experiment.

\subsection{Influence of the flywheels}\label{subsec:EE_flywheel}
As the flywheels play an important role in the control of the motors, this was examined more specifically. Especially, the acceleration levels 10$^{-2}$, 10$^{-3}$ and 10$^{-4}$~\textit{g} have been investigated with respect to the changes of the motion behavior. For each of these acceleration levels, five experiments have been conducted. In \cref{fig:Flywheel}, the envelopes together with the combined mean of the encoder data of the two motors are shown. Those are divided in with and without flywheels based on the adjusted acceleration levels from 1$\cdot$10$^{-2}$ to 10$^{-4}$ \textit{g} (from left to right). The acceleration was determined from the velocity data recorded over time by the encoder. \newline 
It can clearly be seen that the previous rapid, irregular fluctuations and the accelerations' amplitudes of the motors have been smoothed by the flywheels. The flywheels act as low-pass filters for the motors. Although the setup was adjusted for accelerations of 10$^{-2}$~\textit{g}, as previously described, the motors are capable to simulate gravity levels down to 10$^{-3}$~\textit{g}. Furthermore, the amplitudes of these deviations became smaller, leading to a further positive effect generated by the flywheels. \newline 
However, it is also seen that the controller is not perfectly adjusted, causing large overshoots. As a consequence, the controller needs 0.6~s at the beginning to approach its optimum. In connection with this, the influence of the controller decreases as the adjusted gravity level becomes smaller resulting in a better adaptation to smaller accelerations. From the moment the controller reaches its optimum, the deviations based on the hull curve are approximately 1$\cdot$10$^{-3}$~\textit{g} at 10$^{-2}$~\textit{g} regarding the mean plot between 1 and 2~s (in \cref{fig:Flywheel}, bottom left). This is also seen at 10$^{-3}$~\textit{g}. 
Regarding the combined mean, the maximum deviation and thus difference from the ideal value is approximately $\pm5\cdot$10$^{-4}$~\textit{g} at 10$^{-3}$~\textit{g} (\cref{fig:Flywheel}, bottom center). With respect to this, the limits of the current system regarding its adjustable acceleration are also shown, which is highlighted in the plots regarding 10$^{-4}$~\textit{g}, as the amplitudes are too large. Between 1.5 and 2~s, the mean without the flywheels appears improved. However, the envelope shows larger fluctuations. Moreover, the control parameters considering the case with flywheels have not been adjusted. 
\newline In summary, the measurements show the expected improvements by the flywheels. Moreover, they show an improved motor control with maximum deviations of $\pm5\cdot10^{-4}$~\textit{g} at  10$^{-3}$~\textit{g}.
\begin{figure*}[t!]
	\includegraphics[width=15cm]{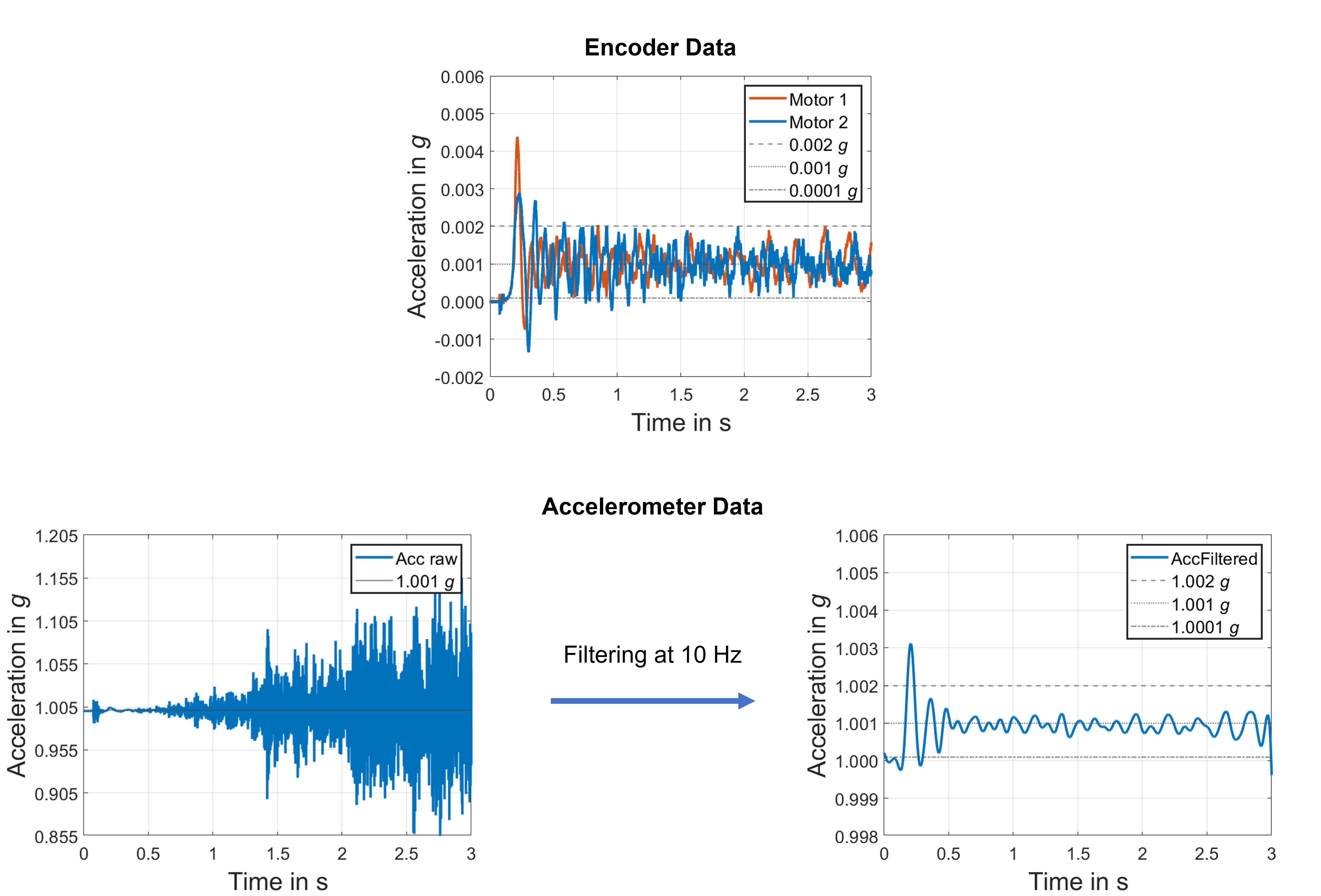}
	\caption{The acceleration sensor was filtered at 10 Hz to show the movement generated by the motors, although this does not apply to the encoder.}
	\label{fig:Filter}
\end{figure*}
\subsection{Comparison of Tests under Earth gravity and Microgravity}\label{subsec:ex_micro}
\begin{figure}[]
	\includegraphics[width=8cm]{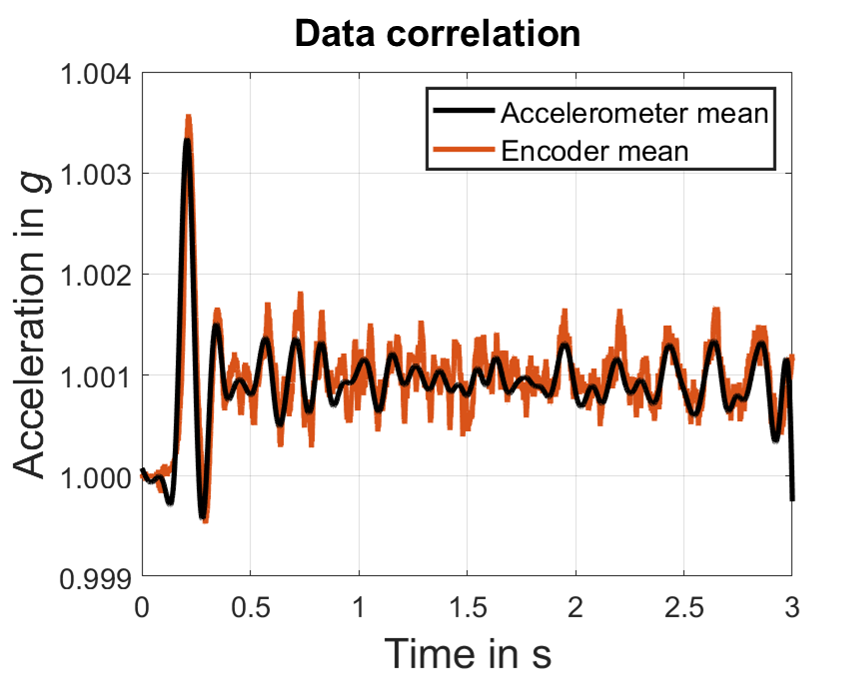}
	\caption{The time series of the accelerometer data at 10 Hz and the data based on the mean of both encoders. The plot shows a strong correlation between the accelerometer data and the mean data of both encoders.} 
	\label{fig:CORR}
\end{figure}

For the evidence inside the EE, for 10$^{-2}$, 10$^{-3}$, and 10$^{-4}$~\textit{g} five measurements have been conducted, giving a total of 15 measurements. Those experiments were performed under atmospheric pressure rather than vacuum due to limited available time, which would have restricted the number of possible test runs. The data from the acceleration sensor of the experiments has been filtered with a low-pass filter at 10 Hz to show the relevant components of the acceleration data (\cref{fig:Filter})\cite{McPherson.2009, Urban.11.12.2015}. \newline 
The reason to apply such a filter is also seen in \cref{tab:coherence} and \cref{fig:CORR}. The coherence values calculated from the unfiltered data show a significant correlation between the accelerometer and the encoders mean at frequencies between 0.1 and 10~Hz. For frequencies above 10 Hz the correlation decreases. Especially, the data for the individual motors show no significant correlation with the accelerometer data. The mentioned correlation is also visible in \cref{fig:CORR}, where the accelerometer data, which has been filtered at 10~Hz, is plotted together with the mean of both encoder data. 
\begin{table}[h]
  \centering
  \caption{Band-averaged coherence between the raw accelerometer and encoder signals for a single flight. The values indicate the degree of coherence in each frequency band. Most of the coherence is found between 1-10 Hz regarding the mean of both encoders $C_\text{mean}$ and the accelerometer signal. In contrast, the individual encoder signals $C_\text{enc1}$ and$C_\text{enc2}$ show much lower coherence in this band, likely due to the vibration-like fluctuations also seen in FIG. 7 (top). At higher frequency bands, moderate coherence is observed regarding the individual encoder signals.}
  \label{tab:coherence}
  \begin{tabular}{l c c c}
    \hline
    Frequency band [Hz] & $C_\text{mean}$ & $C_\text{enc1}$ & $C_\text{enc2}$ \\
    \hline
    0.1  -- 1   & 1.000 & 1.000 & 1.000 \\
    1    -- 10  & 0.992 & 0.197 & 0.199 \\
    10   -- 40  & 0.490 & 0.413 & 0.137 \\
    40   -- 60  & 0.541 & 0.514 & 0.146 \\
    60   -- 100 & 0.520 & 0.485 & 0.204 \\
    \hline
  \end{tabular}
\end{table} Furthermore, the mean acceleration of the unfiltered data is 1.001126~\textit{g} over the 3~s time interval. To investigate the contribution of the high-frequency oscillations to this value, the same signal was high-pass filtered above 40 Hz. Here, the mean acceleration over the same interval is 2.59$\cdot$10$^{-19}$~\textit{g}. This demonstrates that, although high-frequency oscillations are present in the data, they do not contribute to the time-averaged acceleration over the measurement interval.
\begin{figure}[]
	\includegraphics[width=8cm]{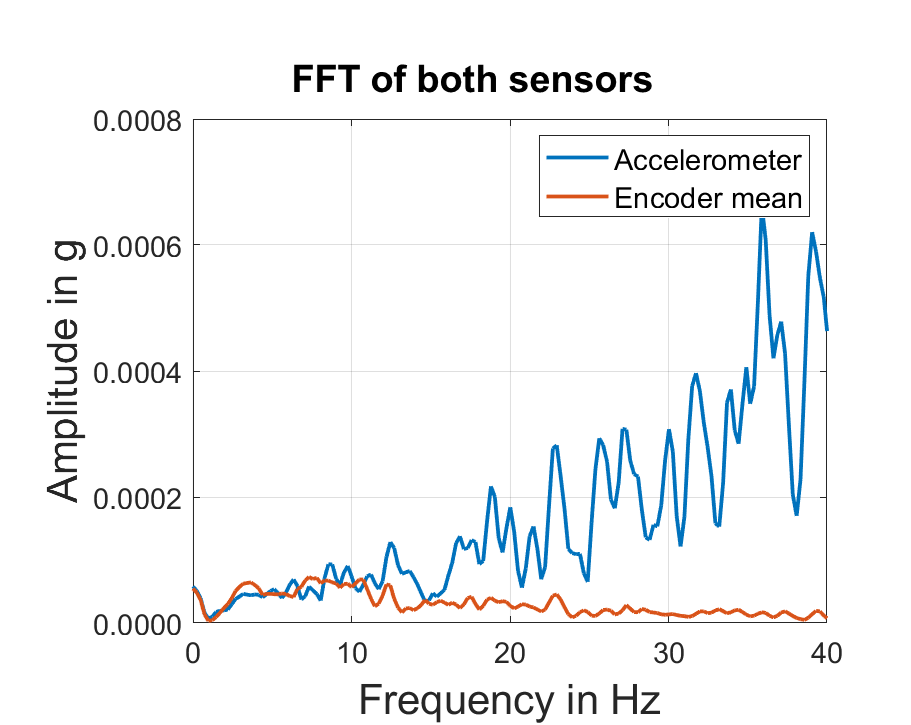}
	\caption{Comparison based on a Fast Fourier Transform (FFT) of both sensor types with  an applied acceleration of 10$^{-3}$~\textit{g} at Earth gravity conditions. It is observed, that the highest amplitudes of the motors' encoders are located between 3 and 13~Hz and that the amplitudes decrease with higher frequencies. The amplitudes of the accelerometer are comparable to those of the encoder around 10~Hz. However, at higher frequencies, they increase.} 
	\label{fig:FFT}
\end{figure} Consequently, these high-frequency components have no significant influence on the experiment's mean acceleration over this interval.\newline In addition, a Fast Fourier Transform (FFT) up to 40 Hz is shown with an applied acceleration of 10$^{-3}$~\textit{g} at Earth's gravity (\cref{fig:FFT}). It can be seen, that the encoder mean and accelerometer show strong coherence between 1 to approximately 13~Hz. Some maxima also correlate at higher frequencies, for example at 23 Hz with a maximum amplitude of approximately $3\cdot10^{-4}$~\textit{g}. However, this is smaller than the targeted acceleration of 10$^{-3}$~\textit{g}. In addition, the accelerometer amplitudes increase progressively starting at 15 Hz, whereas the encoder amplitudes decrease. Nevertheless, it does not exceed $10^{-3}$~\textit{g} and, as calculated above, those frequencies do not contribute to the time-averaged acceleration over the measurement. \newline 
As a consequence, the mean and the envelopes of five experiments, based on the data adjustments, are shown in \cref{fig:comparison} for each gravity level to improve the comparison. Besides the combined mean of the encoders data recorded at microgravity (top, encoder data at Earth's gravity is seen in \cref{fig:Flywheel}), the acceleration sensor has been used additionally to measure the real conditions both with Earth gravity and inside the EE in the microgravity environment (center and bottom). Comparing the encoder with the accelerometer data under Earth gravity, it is seen that those are similar to each other. This was also indicated by the coherence analysis. Especially, the mean of both encoders and the acceleration data correlate again, which was also the case in \cref{fig:CORR}. This is also an indication that the large oscillations of the second motor do not have much influence on the quality of the adjusted acceleration, as also proofed by the coherence analysis.
As described in the previous \cref{subsec:EE_flywheel}, the graphs show the typical behavior of the controller at startup, with the amplitudes becoming smaller as the set acceleration decreases. Furthermore, a good acceleration quality of 10$^{-3}$~\textit{g} is achieved. \newline
\begin{figure*}[]
	\includegraphics[width=19cm]{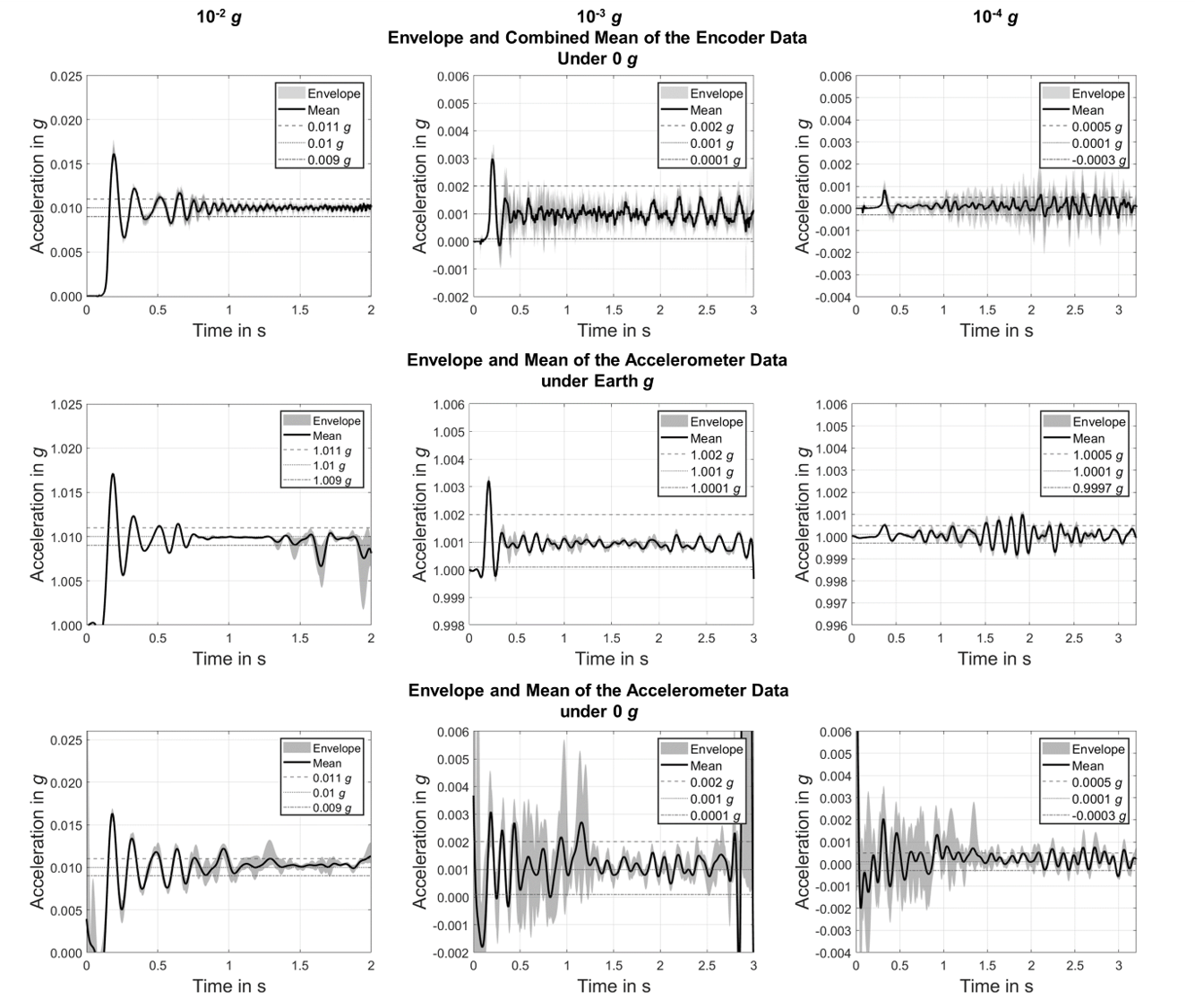}
	\caption{Comparison regarding the sensors and the environment. The accelerations have been recorded by the encoders (top; under microgravity conditions) and the acceleration sensor (center and bottom). Besides, the data of the acceleration sensor is divided in under Earth \textit{g} (center) and microgravity (bottom). The adjusted acceleration are 10$^{-2}$~\textit{g} (left), 10$^{-3}$~\textit{g} (center), and 10$^{-4}$~\textit{g} (right).}
	\label{fig:comparison}
\end{figure*}
Another interesting aspect connected to this is seen from 10$^{-2}$ to 10$^{-3}$~\textit{g} under Earth gravity. Here, the acceleration only differs slightly from the adjusted acceleration similar to the encoder data. However, as previously described and expected, a stable acceleration of 10$^{-4}$~\textit{g} is not detectable with the acceleration sensor, either. Nevertheless, the time range from 0.6~s to 1.5~s shows that a stable control of this gravity level should be possible.\newline
Moreover, the accelerometer graph of 10$^{-2}$~\textit{g} shows two drops at approximately 1.5 ~s (\cref{fig:comparison} center left). This could be an indicator of a suboptimal motion transmission at higher rotational speeds, as the motors' rotational motion are transferred via rotary feedthroughs and metal bellow couplings with clamping hubs to the spindle axes. The envelope even shows a larger deviation. It is conceivable that the couplings used do not maintain sufficient clamping force/contact pressure. The centrifugal loading, caused by the rotational speed and the mass of the rotating components, could overcome the clamping force. However, this is not observed in the experiments conducted in the EE as the resistance force is smaller in microgravity. \newline 
In this context, another crucial aspect is indirectly highlighted, which is the necessity of the acceleration sensor. Without it, any kind of motion transfer errors could not be detected and no clear statements could be made about the acceleration at the examined location since the encoders still show good values regarding their accelerations. Especially at small accelerations, such as in the range of 10$^{-4}$~\textit{g}, the formerly mentioned point is a key factor. This is also seen in the bottom graphs in \cref{fig:comparison} demonstrating the behavior of the experiments inside the EE. In these, several sources of interference can be seen compared to the graph above regarding Earth's gravity at 0 to 1.4 ~s. The reason can be an excitation leading to vibrations and therefore residual accelerations caused by the 5~\textit{g} acceleration phase of the EE. \newline 
Additionally, as the experiments were conducted under normal atmospheric pressure and not vacuum inside the gondola, also an acoustic excitation of the carrier is possible. As initially mentioned, another source of interference can be the imperfect realization of the experimental setup. This includes for instance the motor cables (which are placed outside the vacuum chamber) and other components, which can cause residual vibrations and, consequently, accelerations. In this context, these residual accelerations have to be observed and considered during the conduction of the experiments. Besides all of this and after the excitation phase, the quality of the adjusted acceleration approaches the same level as under Earth's gravity at 10$^{-3}$ and 10$^{-4}$~\textit{g}. As a result, those interference sources do not have a major influence at higher accelerations, such as in the 10$^{-2}$~\textit{g} regime. Consequently, the system shows a good performance at higher gravity levels.
\newline In summary, the results demonstrate on the one hand that an acceleration-based control is necessary to achieve constant accelerations in contrast to an encoder-based control. On the other hand, the results can be considered positive, giving potential for improvements and therefore better results in the future.

\section{Discussion}\label{sec:dis}
The results indicate that the developed acceleration system of AKUS, based mainly on two servo motors and the spindle axes, is in general able to generate the desired accelerations. However, the added flywheels are required for this. As presented before, most of the servo motors are not designed to generate small constant acceleration torques, which is also shown in \cref{fig:OS}. Using the flywheels, the inertia of the whole system becomes higher, resulting in a sufficient motion behavior and therefore constant accelerations. In addition, the motor encoders show a good performance down to an acceleration of 10$^{-3}$~\textit{g}. The maximum deviation here is approximately at 5$\cdot$10$^{-4}$~\textit{g}, although the flywheels were designed for the higher limit, namely  10$^{-2}$~\textit{g}. 
\newline In this context, an adjustment of the system using flywheels with higher inertia can contribute to a better performance in the area of  10$^{-4}$~\textit{g}. Furthermore, the described results have also been reflected by the acceleration sensor under Earth's gravity conditions. Moreover, the mean of the accelerations generated by both motors were similar to those recorded by the acceleration sensor. This indicates that the large amplitudes of the second motor, which are generated to adjust to the first motor, have no negative effects. 
\newline Nevertheless, for higher accelerations starting at 10$^{-3}$~\textit{g}, the controller shows significant transient response with a high first overshoot, indicating that the control parameters have to be adjusted for these accelerations. Improving these parameters will lead to a quicker and an improved steady state condition of the system with smaller deviations. Nevertheless, this behavior is negligibly small at 10$^{-4}$~\textit{g} since the amplitudes occurring at this acceleration level are nearly the same.
\newline The importance of the acceleration sensor is highlighted at accelerations of 10$^{-2}$~\textit{g} tested under Earth gravity and at lower accelerations in microgravity, since it records the actual acceleration at the sample location. For clarity, the accelerometer data are also shown in both raw and low-pass filtered form at 10 Hz, where the filtering is used to highlight the acceleration associated with the system motion. These plots have shown that the sensor is able to detect movement errors or interference sources in the environment underlining its importance. Furthermore, a dynamic control of the acceleration is necessary as the experiments conducted in microgravity indicate. This was also intended from the beginning and is planned in the future.

\section{Conclusion and Outlook}\label{sec:CaO}
Currently, SSSB research primarily depends highly on time and plan-consuming space missions. Although such missions are important for the validation of diverse theories, there have been only few of them. Consequently, it is important to develop Earth-bounded complementary research platforms and laboratories that allow the research of these objects under controlled, realistic conditions. Besides vacuum and low temperature conditions, a key aspect is the enabling of gravity levels between 10$^{-2}$ and 10$^{-4}$~\textit{g}, which are also present on these objects. For this reason, an acceleration system has been developed within the AKUS project to enable such conditions inside the EE, a third generation drop tower. 
\newline The acceleration system is based on two drive trains each consisting of a servo motor and a spindle axis. The system will move a SSSB-simulant on an accelerated trajectory within a vacuum chamber. However, servo motors are not designed to enable such small accelerations as the main purpose of those motors is to perform position- and speed-based tasks. Therefore, adjusted motors with longer shafts and implemented flywheels designed to enable accelerations around 10$^{-2}$~\textit{g} have been added to the motor shafts to increase the system inertia. 
\newline  General tests showed that the flywheels improve the motion behavior as expected. In addition to the encoder-based data, an acceleration sensor placed on the sample holder was used to verify the results besides the encoders. These data confirmed that the motors are capable of simulating the desired acceleration with the adjustments described. The system even achieved acceleration conditions with maximum deviations of approximately $\pm5$$\cdot$10$^{-4}$~\textit{g} at 10$^{-3}$~\textit{g}, which exceeded our expectations. 
\newline For this reason, it can be assumed that the motors can deliver better results when further adjustments are applied, so that the lower limit of 10$^{-4}$~\textit{g} can be achieved in the future. Nevertheless, some mechanical adjustments, such as new couplings or larger flywheels for the smaller accelerations, are necessary to improve the whole system. Subsequent to those adjustments, several tests will be required.
\newline As a result, an overall improvement of the system is one major goal in the future, so that scientific experiments based on the comet-like samples can be conducted. Furthermore, more tests have to be conducted, especially to provide stable accelerations in the full range. In this context, the controller has to be optimized as the current results showed significant overshoots starting at accelerations of 10$^{-3}$~\textit{g} and higher. Moreover, alternative control concepts have to be developed, such as a cascade control based on sensor fusion using the encoders and the acceleration sensor. Especially, the implementation of the acceleration sensor for control is an important step to improve the system. This has to be considered with respect to the residual acceleration of the EE at the beginning of the microgravity time.
\newline In this context, firstly tests in vacuum and finally test with the comet-like sample will be performed in the near future as this is the main goal of the project AKUS. As a result, the entire system has to be built up including the cameras, which are necessary to observe the sample depending on the acceleration. The setup can then be used for the study of SSSB (such as comets and asteroids), providing an Earth-bounded complementary research opportunity leading to international research projects.

\begin{acknowledgments}
The authors would like to thank the German Space Agency within the German Aerospace Center (DLR) and the Federal Ministry for Economic Affairs and Climate Action (BMWK) on the basis of a decision by the German Bundestag (FKZ: 50WM2254). Moreover, we would like to thank the DFG  and the Lower Saxony state government for their ﬁnancial support for building the Hannover Institute of Technology (HITec) and the Einstein-Elevator (NI1450004, INST 187/624-1 FUGB) as well as the Institute for Satellite Geodesy and Inertial Sensing of the German Aerospace Center (DLR-SI) for the development and the provision of the experiment carrier system. Further funding was provided by the D-A-CH program (DFG: GU 1620/3-1 and BL 298/26-1, project number 395699456; SNF: 200021E 177964; FWF: I 3730-N36) of the CoPhyLab, the European Union under grant agreement NO. 101081937 – Horizon 2022 - Space Science and Exploration Technologies (views and opinions expressed are however those of the authors only and do not necessarily reflect those of the European Union; neither the European Union nor the granting authority can be held responsible for them) and the Niedersächsisches Vorab in the framework of the research cooperation between Israel and Lower Saxony under grant ZN 3630.

We would also like to thank Jan Hoffmann and Mohammad Ibrahim for their contributions during their student work.

\end{acknowledgments}

\section*{AUTHOR DECLARATIONS}

\subsection*{Conflict of Interest}
The authors declare no conflicts of interest.

\subsection*{Special Unit}
\textit{g} = Earth's gravity ($1\,\textit{g} = 9.81~\text{m/s}^2$)

\section*{Data Availability Statement}
The datasets used and/or analysed during the current study are available from the corresponding author or the co-authors on reasonable request.

\nocite{*}
\bibliography{AKUS_ref}

\end{document}